\documentclass{iaus}
\usepackage{graphicx}

\title[PTPS stars radii determination with CHARA Array] 
{PTPS candidate exoplanet host star radii determination with CHARA Array}

\author[P. Zieli\'nski \etal]   
{Pawe\l{} Zieli\'nski$^{1}$, Martin Va\v{n}ko$^{2}$, Ellyn Baines$^{3}$, Andrzej Niedzielski$^{1}$ \and  Aleksander Wolszczan$^{4}$
 }

\affiliation{$^1$ Toru\'n Centre for Astronomy, Nicolaus Copernicus University, Gagarina 11, 87100 Toru\'n, Poland\\ 
										email: {\tt pawziel@astri.umk.pl} \\[\affilskip]
   $^2$ Astronomical Institute, Slovak Academy of Sciences, 05960 Tatransk\'a Lomnica, Slovakia\\[\affilskip]
   $^3$ Naval Research Laboratory, Remote Sensing Division, 4555 Overlook Ave. S.W., Washington, DC 20375\\[\affilskip]
   $^4$ Department for Astronomy and Astrophysics \& Center for Exoplanets and Habitable Worlds, Pennsylvania State University, 525 Davey Laboratory, 		University Park, PA 16802}

\pubyear{2011}
\volume{282}  
\jname{From Interacting Binaries to Exoplanets: Essential Modelling Tools}
\editors{A.C. Editor, B.D. Editor \& C.E. Editor, eds.}
\begin{document}

\maketitle

\begin{abstract}
We propose to measure the radii of the Penn State - Toru\'n Planet Search (PTPS) exoplanet host star candidates using the CHARA Array. Stellar radii estimated from spectroscopic analysis are usually inaccurate due to indirect nature of the method and strong evolutionary model dependency. Also the so-called degeneracy of stellar evolutionary tracks due to convergence of many tracks in the giant branch decreases the precision of such estimates. However, the radius of a star is a critical parameter for the calculation of stellar luminosity and mass, which are often not well known especially for giants. With well determined effective temperature (from spectroscopy) and radius the luminosity may be calculated precisely. In turn also stellar mass may be estimated much more precisely. Therefore, direct radii measurements increase precision in the determination of planetary candidates masses and the surface temperatures of the planets.
\end{abstract}

\firstsection 
\section{Motivation}
Within the Penn State - Toru\'n Planet Search (PTPS, \cite[Niedzielski \& Wolszczan 2008]{NW2008}), which is a radial velocity project to search for and characterize planets around stars more massive than the Sun, the stellar integral properties ($M$, $R$ and $L$) are currently determined from a combination of atmospheric parameters (surface gravity, $T_{\rm{eff}}$ and [Fe/H]) together with the evolutionary tracks (\cite[Girardi \etal\ 2000]{G2000}). Unfortunately this method is uncertain due to difficulties in accurate placement of individual stars on the H-R diagram.

\begin{figure}[h]
\begin{center}
 \includegraphics[angle=-90,width=9cm]{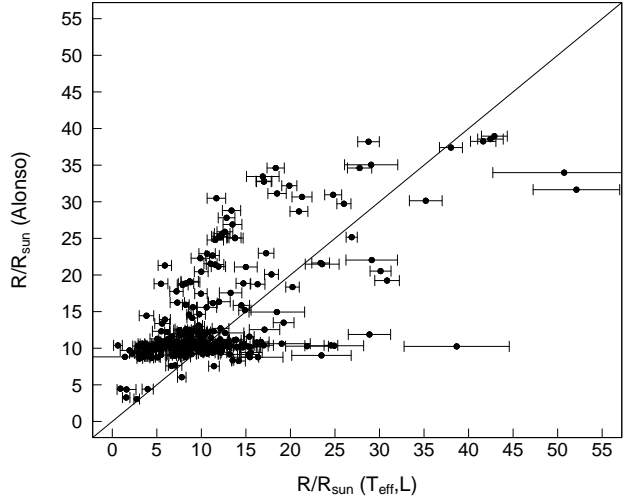} 
 \caption{The comparison of the $R/R_{\rm{\odot}}$ obtained in the spectroscopic study of PTPS subsample of 332 stars and from the empirical calibration of 
\cite[Alonso \etal\ (2000)]{A2000}. Only the uncertainties in PTPS determinations are shown. The solid line denotes the one to one relation.}
   \label{fig1}
\end{center}
\end{figure}

Fig.~\ref{fig1} presents the comparison between the determinations of stellar radii from our spectroscopic analysis (\cite[Zieli\'nski \etal\ 2011]{Z2011}) and the empirical calibration by \cite[Alonso \etal\ (2000)]{A2000}. Both results are in general agreement, nevertheless, large discrepancies for individual objects are visible. The uncertainties of our radii determinations based on stellar $T_{\rm{eff}}$ and $L$ was found to be 0.8~$R_{\rm{\odot}}$ on average but due to missing parallaxes the results  are strongly dependent on adopted stellar evolutionary models as well as on photometric data quality. We plan to improve this by using the CHARA Array (\cite[ten~Brummelaar \etal\ 2005]{B2005}) for direct measurements of stellar radii and, in turn,  using the relation $g = \gamma M R^{-2}$, constrain better the stellar mass. On the other hand, the luminosity  can be calculated with high accuracy when using the Stefan-Boltzmann law directly.

Two stars from PTPS survey were already observed with the CHARA Array: HD~17092 (\cite[Baines \etal\ 2009]{B2009}) and HD~214868 (\cite[Baines \etal\ 2010]{B2010}). The results obtained from direct measurements agree well with the spectroscopic values estimated by us (see \cite[Niedzielski \etal\ 2007]{N2007}, \cite[Zieli\'nski \etal\ 2011]{Z2011}). However, the accuracy  of interferometry-based measurements is significantly better.

\section{Targets selection}
Our observing list was derived from the PTPS survey of evolved stars with RV planet candidates and comprises 212 stars suitable for CHARA observations. These stars have already been observed spectroscopically using the Hobby-Eberly Telescope (\cite[Ramsey \etal\ 1998]{R98}), so high-resolution spectra are available. The targets are sufficiently bright (V $ < 8.5$~mag and K $ < 6$~mag) and nearby ($\pi \ge 4.03$~mas) giants with large radii that can be easily directly measured interferometrically.

\section{Expected results}
Using new CHARA measurements we will be able to obtain high-precision radii measurements and in turn better estimates of the stellar masses. With the angular diameter of a precision of 2 \%, the stellar radius can be determined with a precision better than 5 \%. The largest contributor to the error budget is the parallax, therefore, after taking into account  10 \% parallax uncertainty, the final mass will then be precise within 0.1~$M_{\rm{\odot}}$. This is significantly better compared to our current estimates based on moderate quality photometry and evolutionary models. For  PTPS targets studied in detail, we estimated masses between 1 and 3~$M_{\rm{\odot}}$ and even after critical assessment an uncertainty of 30 \% remains. Hence, the CHARA data will improve the precision in mass sufficiently for the subsequent analysis.

\begin{acknowledgements} 
The research leading to these results has received funding from the European Community's Seventh Framework Programme under Grant Agreement 226604. PZ and AN acknowledge the financial support from the Polish Ministry of Science and Higher Education through grant N N203 510938. MV is supported by VEGA 2/0094/11. AW acknowledges support from NASA grant NNX09AB36G. The Center for Exoplanets and Habitable Worlds is supported by the Pennsylvania State University, the Eberly College of Science, and the Pennsylvania Space Grant Consortium.
\end{acknowledgements}

\end{document}